\documentstyle[sw20aip]{article}

\input{tcilatex}

\begin{document}

\title{Aging and L\'{e}vy distributions in sandpiles.}
\author{Oscar Sotolongo-Costa,$^{1,2}$ A. Vazquez$^{2}$ and J.C. Antoranz$^{1}$ \\
$^{1}$Depto. F\'{\i}sica-Matem\'{a}tica y Fluidos, Fac. Ciencias. \\
UNED, 28080 Madrid\\
$^{2}$Depto. de F\'{\i}sica Te\'{o}rica, Fac. F\'{\i}sica, Univ. Habana. \\
Habana 10400, Cuba.}
\date{\today}
\maketitle

\begin{abstract}
Aging in complex systems is studied ${\it via}$ the sandpile model.
Relaxation of avalanches in sandpiles is observed to depend on the time
elapsed since the beginning of the relaxation. L\'{e}vy behavior is observed
in the distribution of characteristic times. In this way, aging and
self-organized criticality appear to be closely related.
\end{abstract}

\section{\protect\smallskip Introduction}

In the current literature of complex systems, aging is frequently reported
as a phenomenon observed in spin glasses and other disordered systems,
rather than in biological systems \cite{1,2,3,4,5,6}, though some works have
been devoted to biological aging of populations \cite{7,8}. Relevant
contributions have been brought presenting this phenomenon from a general
perspective, revealing that temporal autocorrelation functions in the
Bak-Sneppen model display aging behavior similar to glassy systems\cite{9,10}%
. To our knowledge those papers give, up to now, the more relevant
contribution to the relation between SOC and aging.

Up to our knowledge, a detailed study of individual aging from the viewpoint
of self-organized critical systems has not yet been performed.

It is well known that biological aging manifests itself in an individual as
the slowing down of many processes as, {\em e.g}., slower growing of tumors
and manifestation of senescence in the slowing down of reflex behavior. Due
to its characteristics, relaxation phenomena in complex systems are good
candidates to the study of aging.

About 150 years ago, Weber and Gauss carried out a simple experiment
demonstrating that relaxation in complex systems is not exponential.
Investigating the contraction of a silk thread they found that it does not
contract so quickly but it relaxes slowly following a power law $t^{-\alpha
} $. This behavior is not particular of mechanical relaxation, it has been
also observed in experiments of magnetic relaxation in spin glasses and high
critical temperature superconductors, transient current measurements in
amorphous semiconductors, dielectric relaxation, and more \cite{rel}.

In relaxation dynamics, aging means that the properties of a system depends
on its age \cite{bou}. For instance, consider a glass quenched at time $t=0$
below its glass transition temperature under an external stress. At time $%
t=t_w$ the stress is released. If the system were near equilibrium, then its
response measured at certain time $t>t_w$ will be a function of the
difference $t-t_w$ and, therefore, normalized responses taken at different
initial times $t_w$ will collapse in a single curve. However, this is not
the behavior in glass materials, and, as was pointed above, in live beings.

Aging is a consequence of the nonequilibrium dynamics and may be considered
as a characteristic of the dynamics of complex systems far from equilibrium.

In the following sections we demonstrate that these manifestations of
individual aging are present in sandpile avalanches modeled with a Bethe
lattice, particularly, the process of relaxation. Contrary to \cite{9,10},
we do not consider deterministic ''toppling'' (like that present in the
Bak-Sneppen model) but a probabilistic one, and avalanches are not regarded
as infinite, as is \cite{9}, but the finite size of the system (sandpile) is
explicitly included.

\section{Sandpile model}

Sandpiles seem to be simplest systems which lead to complex behavior and are
a paradigm for the study of all phenomena manifested in complex systems out
of equilibrium, like relaxation under external perturbations. They have been
taken as the paradigm of self-organization since the introduction of these
ideas by Bak, Tang, and Wiesenfeld \cite{bak}. The evolution of a sandpile
under an external perturbation has been extensively studied experimentally 
\cite{jae,hel}, theoretically \cite{tan,zap}, and by computer simulations 
\cite{kad}. In general, those works analyze the sandpile dynamics close to
the critical state, i.e., close to the critical angle.

If a sandpile is subjected to an external perturbation, like, {\em e.g.},
small amplitude vibrations, large amplitude grain motions are rare events
but from time to time, a grain may jump from its quasiequilibrium position.
If a surface grain jumps, it will fall through the slope of the pile until
it collides with another sand grains down the slope. After this collision
the initial grain may be trapped with those grains or some of them may fall
through the slope. If the last possibility happens then each grain
''surviving'' the collision will fall through the slope as the initial grain
did. This image of an avalanche as an initial object that consecutively
drags another resembles a branching process for which the Bethe lattice
representation seems to be natural \cite{zap,osc}. The representation of the
avalanche dynamics as a Bethe lattice is a mean field approach to the
problem. It neglects correlations between branches. Notwithstanding, if the
pile angle is below is critical value the avalanches will be rare events
and, in case of occurrence, will be very sparse. Thus, such an approximation
will be acceptable for analyzing the long time relaxation of the pile angle
below its critical value, which is the subject of this paper. Away from the
critical point the mean field approximation works quite well.

Let us represent the avalanche as a cascade in the Bethe lattice as follows.
Firstly, we start with a single node, which could represent in this case a
grain. In a further step $F$ will emerge with probability $p(\theta )$,
depending on the pile angle $\theta $. This operation of generation of $F$
identical particles starting from one is repeated in the next step to each
node of the new group, and so on. If the percolation process overcomes a
given length (that of the border of the pile) those nodes beyond the limit
constitute the avalanche. If it does not, there would be no avalanche since
the cascade was stopped before reaching the base of the pile (frustrated
avalanche). By avalanche size we take the number of nodes, of the
corresponding Bethe lattice, in the last step.

After an avalanche the sandpile {\em autoorganizes itself} with the new
number of grains (i.e., a new slope is calculated with the remaining
grains). The occurrence of avalanches will carry as consequence a decrease
in the number of grains in the pile and, therefore, a decrease of $p(\theta )
$. Each time an avalanche occurs, the occurrence of a new avalanche is less
probable. Thus, to characterize this feedback mechanism a relation between $p
$ and the number of grains in the pile $N$ is needed.

The drag probability $p(\theta )$ is a function of the slope. Its value is
determined by the competition of two contrary forces: the gravity, which
conspires against the stability of the slope, and the friction which favors
the slope stability. Since the slope forms an angle $\theta $ with the
horizontal plane the component of the gravity force in the slope direction
will be larger with the increase of this angle, varying from zero to a
maximum value when $\theta $ goes from zero to $\pi /2$. Therefore it is
plausible to assume that the contribution of the gravity and, therefore, the
tendency of falling down the slope is proportional to $\sin \theta $. On the
contrary, the resistance to this tendency given by the static friction
decreases with decreasing the pile angle, varying from a maximum value to
zero when $\theta $ goes from zero to $\pi /2$. Hence, it is also plausible
to assume that the resistance to the falling down is proportional to $\cos
\theta $.

Thus the slope dependence of $p$ can be expressed through the ratio of both
tendencies $\sin \theta /\cos \theta =\tan \theta $. Based on this
hypothesis we may propose the exponential relation 
\begin{equation}
p(\theta )=\exp (-A/\tan \theta )\ ,  \label{eq:1}
\end{equation}
where $A$ is a parameter determined by the gravitational field, the
friction, and vibration intensity. Notice that $p(0)=0$ and $p(\pi /2)=1$.
This selection for $p(\theta )$ can not be regarded as a sophisticated trick
or something imposed {\em ad hoc }to the model in order to obtain the
desired results, since it can be verified that {\em any }function satisfying
the very general conditions already stated is adequate to our model.
Incidentally, it must be said that this reveals the robustness of the model.

On the other hand, the number of grains in a pile with slope angle $\theta $
is given by 
\begin{equation}
N=Bx^3\tan \theta ,  \label{eq:2}
\end{equation}
where $B$ is a geometrical factor and $x=D/d$ is the ration between a
characteristic size of the pile base $D$ and sand grain $d$. Combining
equations (\ref{eq:1}) and (\ref{eq:2}) it is obtained

\begin{equation}
p(N)=\exp (-cx^3/N)\ .  \label{eq:3}
\end{equation}
where $c=AB$. Thus Equation \ref{eq:1} relates the dragging probability with
the number of grains in the pile.

The parameter $c$ should not depend on the size of the system. It must be a
function of parameters describing the vibrations, amplitude $a$ and
frequency $\omega $, and of the gravitational field acceleration $g$. The
only nondimensional combination of these magnitudes is given by the ratio
between the vibration acceleration $a\omega ^2$ and $g$ and, therefore, $%
c=c(a\omega ^2/g)$. This conclusion, obtained from dimensional analysis, is
corroborated by experiments on sandpiles under vibrations \cite{eve} which
show that the ratio $a\omega ^2/g$ is the relevant parameter.

For a given value of $c$ and $x$ , there is a critical number of grains in
the pile $N_c$. This value can be found recalling that in the Bethe lattice
the critical value of $p$ for percolation exists is $1/F$, resulting 
\begin{equation}
N_c=cx^3/\ln F\ .  \label{eq:4}
\end{equation}
In a static sandpile (i.e., no vibrations) this value corresponds with the
number of grains in the pile at the angle of repose~\cite{zap}.

Another magnitude of interest is the penetration length, the number of steps 
$n(\theta )$ in the Bethe lattice for an avalanche to take place. This
magnitude must be proportional to the length of the pile slope and,
therefore, should be given by the expression 
\begin{equation}
n(\theta )=x/\cos \theta  \label{eq:4a}
\end{equation}
Notice that the possible existence of a geometrical factor in equation \ref
{eq:4a} may be absorbed in $x$, redefining $B$ in Equation \ref{eq:2}.
Moreover, the relation between $n$ and $N$ may be easily obtained using
Equation \ref{eq:2}.

Equations \ref{eq:3} and \ref{eq:4a} relate the parameters of the pile with
those of its Bethe lattice representation. They were obtained here in a
different way than that proposed by Zaperi {\em et al }\cite{zap}. Other
dependencies between the dragging probability and the number of grains in
the sandpile may be proposed. Notwithstanding, as it is discussed below, the
precise form of this functional dependence is not relevant.

\section{Simulations and results}

\FRAME{ftbpFUw}{2.4872in}{2.5771in}{0pt}{\Qcb{Aging. Normalized relaxation
function $\protect\phi (t+t_{w})=N(t+t_{w})/N(t_{w})$ taken at three
different initial times $t_{w}$. The long time tail exhibited in the figure
has an exponent $\protect\alpha \simeq 0.089.$}}{}{figure1.wmf}{\special%
{language "Scientific Word";type "GRAPHIC";maintain-aspect-ratio
TRUE;display "USEDEF";valid_file "F";width 2.4872in;height 2.5771in;depth
0pt;original-width 5.4267in;original-height 5.6247in;cropleft "0";croptop
"1";cropright "1";cropbottom "0";filename 'figure1.wmf';file-properties
"XNPEU";}}

\bigskip

The numerical experiment of relaxation is performed as follows. We start
with a certain number of $N$. In a first step, we test if an avalanche takes
place using the Bethe lattice representation. If it does, then we simulate
the process and then recalculate the value of $p(N)$ by simply substracting
the size of the avalanche to $N$ and using Equation \ref{eq:3}. The size of
the avalanche is the number of nodes generated in the Bethe lattice which
surpass the size of the pile whose measure is given by (\ref{eq:4a}). Then
this step is repeated again and again, whereas avalanche sizes and times are
registered.

If avalanches are considered as instantaneous the number of steps is a
measure of time. This approximation is valid for low vibration intensities.
In this case grain jumps which trigger avalanches are rare events and,
therefore, the time between two successive grain jumps will be much larger
than the duration of avalanches.

To describe the behavior of our system, we define the relaxation function $%
\phi (t+t_w,t_w)$ of the sandpile as:

\begin{equation}
\phi (t+t_{w,}t_w)=\frac{N(t+t_w)}{N(t_w)},  \label{eq:4b}
\end{equation}

where we include the time dependence of the number of grains in the
sandpile. $t_w$ is the ''waiting time'', i.e., the instant at which we begin
the counting of the number of grains since the start of the relaxation.
(avalanche), as in \cite{9,10}.

The existence of aging in our model is illustrated in Figure 1. We have
plotted the normalized relaxation $\phi (t+t_{w},t_{w})$ at different ages
(steps) of the system (simulation) by taking as initial time $t=t_{w}$ for $%
x=10$ and $c=1$. As it can be seen the relaxation becomes slower with
increasing the age of the system, in agreement with experiments in
structural and spin glasses \cite{bou} (and, unfortunately, with human
life). The system exhibits a delay in the relaxation, after which the
relaxation function decreases as a power law (long time tail) with an
exponent $\alpha \simeq 0.089.$ Thus, since the angle relaxation becomes
slower with the age of the pile then it will never reach an equilibrium
angle and, therefore, properties like translational invariance and the
fluctuation dissipation theorem do not hold \cite{bou}.

\FRAME{ftbpFU}{2.4682in}{2.5763in}{0pt}{\Qcb{Distribution of time between
avalanches. The dashed line is a power tail with exponent $-1-\protect\alpha 
$ , $\protect\alpha $ being the exponent of the long time tail shown in
Figure 1.}}{}{figure2.wmf}{\special{language "Scientific Word";type
"GRAPHIC";maintain-aspect-ratio TRUE;display "USEDEF";valid_file "F";width
2.4682in;height 2.5763in;depth 0pt;original-width 5.3748in;original-height
5.6144in;cropleft "0";croptop "1";cropright "1";cropbottom "0";filename
'figure2.wmf';file-properties "XNPEU";}}

\bigskip

Associated with these slow relaxation dynamics and aging phenomena we expect
to observe a wide distribution of time between avalanches. As time increases
the time between two consecutive avalanches $\Delta t$ becomes larger,
because after an avalanche the occurrence of a new avalanche becomes
smaller. Therefore, it is expected that the mean time between avalanches
diverges as $t\rightarrow \infty $. Hence, the distribution of time between
avalanches $n(\Delta t)$ should satisfy the asymptotic behavior for large $%
\Delta t$ 
\begin{equation}
n(\Delta t)\sim \ \Delta t^{-1-\beta },  \label{eq:5}
\end{equation}
with $0<\beta <1$.

This hypothesis is confirmed in our simulations. Figure 2 shows the
distribution of time between avalanches for $x=10$ and $c=1$. It is
approximately constant for small values and then it decays following a power
law in more than two decades. The plateau at small values of $\Delta t$ is
associated to the rapid decay observed at short times while the tail for
large $\Delta t$ should be related to the long time tail. Here we observe
that $\beta \simeq \alpha \simeq 0.089$ Thus, there should be some
connection between the distribution of time between avalanches and the long
time relaxation.

Finally we want to mention that these simulations were also carried out
assuming other functional dependences between the dragging probability and
the number of grains in the pile $p(N)$. In all cases the results where
qualitatively similar to those presented here using equation \ref{eq:3},
reflecting that the precise dependence is not relevant.

\section{Conclusions}

A sandpile model for aging was presented revealing that this phenomenon is
manifested in relaxation of sandpile avalanches. Biological systems and
individuals show similar behavior. Phenomena related to some kind of
relaxation in live beings deserve more quantitative research. The
introduction of a Bethe lattice representation for the avalanches and a
feedback mechanism describes quite well the principal features of the
relaxation in sandpiles under low intensity vibrations. The proposed
representation leads to long time tails relaxation, aging and L\'{e}vy
(fractal) distributions of time constants, which are characteristic
properties of the dynamics of complex systems out of equilibrium.

\subsection{Acknowledgments:}

This work was partially supported by the {\it Alma Mater} prize, given by
The University of Havana, and the Direcci\'{o}n General de Investigaci\'{o}n
Cient\'{\i}fica y T\'{e}cnica DGICYT, Spain, Ministerio de Educaci\'{o}n y
Cultura.


\begin{thebibliography}{99}
\bibitem{1}  K. Binder, A. P. Young, Rev. Mod. Phys{\bf \ 58}, 801 (1986).

\bibitem{2}  J. Jackle, Rep. Prog. Phys. {\bf 49}, 171 (1986).

\bibitem{3}  B. Derrida, H. Spohn, J. Stat. Phys. {\bf 51}, 817 (1988).

\bibitem{4}  M. M\'{e}zard, G. Parisi, J. Phys. A {\bf 23}, L1229 (1990).

\bibitem{5}  S. John, T. C. Lubenski Phys. Rev. B {\bf 34}, 4815 (1986).

\bibitem{6}  G. Kriza, G. Mah\'{a}ly Phys. Rev. Lett {\bf 56}, 2529 (1986).

\bibitem{7}  T. J. P. Penna J. Stat. Phys. {\bf 78}, 1629 (1995), Preprint
cond-mat/{\bf 9503099}

\bibitem{8}  R. M. C. de Almeida, C. Moukarzel Physica A {\bf 257}, 10
(1998).

\bibitem{9}  S. Boettcher, M. Paczuski Phys. Rev. Lett {\bf 79}, 889 (1997).

\bibitem{10}  S. Boettcher, Phys. Rev. E {\bf 56}, 6466 (1997).

\bibitem{rel}  {\it Relaxation in Complex Systems and Related Topics},
edited by I. A. Cambell and C. Giovanella, NATO ASI series B222 (Plenum
Press, New York, 1990).

\bibitem{bou}  J. P. Bouchaud, L. F. Cugliandolo, J. Kurchan, and M.
M\'{e}zard, {\it Out of Equilibrium Dynamics in Spin-Glasses and Other
Glassy Systems}, cond-mat/{\bf 9702070}.

\bibitem{bak}  P. Bak, C. Tang, and K. Wiesenfeld, Phys. Rev. Lett. {\bf 59}%
, 381 (1987).

\bibitem{jae}  H. M. Jaeger, C. Liu, and S. R. Nagel, Phys. Rev. Lett. {\bf %
62}, 40 (1989).

\bibitem{hel}  G. A. Held, D. H. Solina, D. T. Keane, W. J. Haag, and G.
Grinstein, Phys. Rev. Lett. {\bf 65}, 1120 (1990).

\bibitem{tan}  C. Tang and P. Bak, J. Stat. Phys. {\bf 51}, 797 (1988).

\bibitem{zap}  S. Zapperi, K. B. Lauritsen, and H. E. Stanley, Phys. Rev.
Lett. {\bf 75}, 4071 (1995); K. B. Lauritsen, S. Zapperi, and H. E. Stanley,
Phys. Rev. E {\bf 54}, 2483 (1996).

\bibitem{kad}  L. P. Kadanoff, S. R. Nagel, L. Wu, and S. Zhou, Phys. Rev A 
{\bf 39}, 6524 (1989).

\bibitem{osc}  O. Sotolongo-Costa, A. Vazquez, J. C. Antoranz, Preprint
cond-mat/{\bf 9806176.}

\bibitem{bar}  G. C. Barker and A. Metha, Phys. Rev. E {\bf 47}, 184 (1993);

\bibitem{eve}  P. Evesque and J. Rajchenbach, Phys. Rev. Lett {\bf 62}, 44
(1989).\newpage
\end{thebibliography}
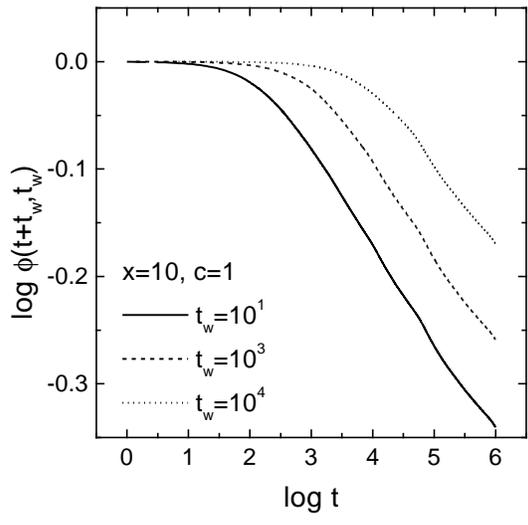
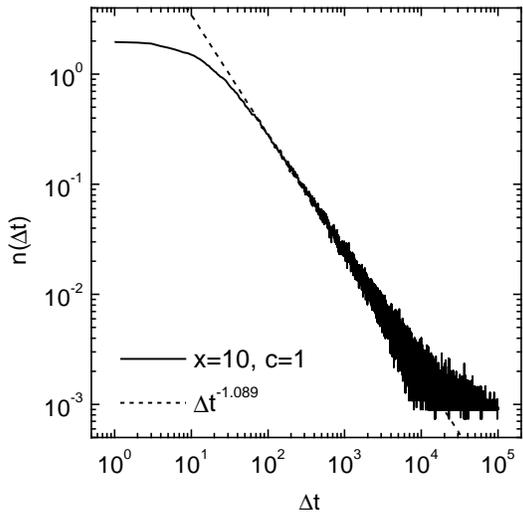
\end{document}